\documentstyle[12pt,epsf]{article}
\newcommand{\be}{\begin{eqnarray}}
\newcommand{\ba}{\begin{array}}
\newcommand{\ea}{\end{array}}
\newcommand{\ee}{\end{eqnarray}}

\newcommand{\cL}{{\cal L}}
\newcommand{\eijk}{{\epsilon_{ijk}}}
\newcommand{\beq}{\begin{equation}}
\newcommand{\eeq}{\end{equation}}
\newcommand{\beqa}{\begin{eqnarray}}
\newcommand{\eeqa}{\end{eqnarray}}

\newcommand{\vs}{\vspace{-0.15cm}}

\begin{document}

\noindent {\bf REVISED VERSION, \today}

\hfill FZJ-IKP(TH)-1998-16

\hfill UNITU-THEP-{12}/1998


\vspace{1in}

\begin{center}

{{\Large\bf The parity--violating pion--nucleon coupling constant\\[0.3em]
   from a realistic three flavor Skyrme model}}

\end{center}

\vspace{.3in}

\begin{center}
{\large
Ulf-G. Mei{\ss}ner$^\ddagger$\footnote{email: Ulf-G.Meissner@fz-juelich.de},
Herbert Weigel$^\dagger$\footnote{Heisenberg--Fellow}\footnote{
Present address: Massachusetts Institute of Technology,
77 Massachusetts Ave, Cambridge, Ma 02139}\footnote{email:
weigel@ctp.mit.edu}
}

\bigskip

$^\ddagger${\it Forschungszentrum J\"ulich, Institut f\"ur Kernphysik
(Theorie)\\ D-52425 J\"ulich, Germany}

\bigskip

$^\dagger${\it Universit\"at T\"ubingen, Institut f{\"u}r Theoretische Physik\\
Auf der Morgenstelle 14, D-72076 T\"ubingen, Germany}\\

\bigskip

\end{center}

\vspace{.9in}

\thispagestyle{empty}

\begin{abstract}\noindent
We study the parity--violating pion--nucleon coupling $G_\pi$ in the
framework of a realistic three flavor Skryme model. We find a
sizeable  enhancement of $G_\pi \simeq 0.8 \ldots 1.3 \cdot 10^{-7}$
compared to previous calculations in two--flavor models with vector
mesons. This  strangeness
enhancement stems from induced kaon fields of the chiral soliton
and  the non-monotoneous dependence on symmetry breaking of the
nucleon matrix element of the flavor singlet piece of the operator
associated with these induced fields. Both features
are sensitive to four quark operators including an $\bar s s$ pair.
\end{abstract}

\vfill

\pagebreak

\noindent {\bf 1.} There has been considerable interest and controversy
about the parity--{violating} pion--nucleon coupling constant
$G_\pi$\footnote{We reserve the symbol $f_\pi$ for the weak pion decay
constant. In many papers, $G_\pi$ is denoted by $h_\pi^1$.} over the
last years, triggered on one side by  new experimental results and
on the other by fresh theoretical ideas. The measurement of the anapole moment
in $^{133}$Cs~\cite{Cs},
which allows to get a bound on $G_\pi$, seems to
contradict the bounds from the anapole moment  measured in $^{205}$Tl~\cite{Tl}
and the bound from the circular polarization asymmetry measurement
of $^{18}$Fl~\cite{Fl}.
To be precise, these data are analyzed in the framework of
parity--violating (pv)
one--boson exchange and are thus sensitive to the products of the weak and
the strong (parity--conserving) couplings. The latter are, however,
sufficiently well known
for the present accuracies one is dealing with. A recent paper which
deals with this topic is ref.~\cite{WB}. On the theoretical side, the
chiral perturbation theory analysis of Kaplan and Savage~\cite{KS} seems
to indicate a large enhancement of the weak pion--nucleon coupling due
to strangeness. More precisely, the underlying four--fermion current--current
Hamiltonian has a piece of the form $(\bar q q)(\bar s s)$, which contributes
{sizeably} to $G_\pi$. Here, $q$ ($s$) denotes the light (strange) quarks.
In ref.~\cite{KS}, numerical estimates {were} given
based on factorization and dimensional analysis. On the other hand, the
two--flavor topological chiral soliton model {was} used to study
parity--violating meson--nucleon couplings~\cite{KM1,KM2} and interaction
regions~\cite{KM3}, including also the $N \Delta$ and $\Delta\Delta$ vertices.
This model gives a successful description of many nucleon observables
as reviewed in ref.~\cite{UGM}.
In this approach, the weak pion--nucleon coupling comes out to be very
small, typically $G_\pi \simeq 0.3\cdot 10^{-7}$. Note that in the
SU(2) Skyrme model without vector mesons, the weak pion--nucleon
coupling vanishes due to a particular symmetry between the
currents~\cite{KM1}. This symmetry is absent in the presence of vector
mesons or realistic three flavor version (we thus always compare to
the vector meson stabilized Skyrmion when we talk about SU(2)).
Clearly, in the
two--flavor approch one is not sensitive to operators involving strange
quarks and also strange components in the nucleon wave function. It appears
therefore mandatory to extend the soliton model calculations to the
three flavor case. In this letter, we calculate $G_\pi$ in a realistic
SU(3) Skyrme model, which gives a fair description of many observables,
like the mass splittings  of the low-lying $\frac{1}{2}^+$ and
$\frac{3}{2}^+$ baryons, magnetic moments, hyperon deacys
and many others (for a review, see ref.~\cite{HW}). Within this
approach, we can quantify the role of the four--quark operators
involving strange quark pairs as well as the role of strangeness in
the nucleons' wave function.

\medskip

 \noindent {\bf 2.}
The three flavor Skryme model is defined by the Lagrange density
\be \label{eq:LS}
\cL&=&-\frac{f_\pi^2}{4}{\rm tr}\
\left(\alpha_\mu\alpha^\mu\right)
+\frac{1}{32e^2} {\rm tr}
\left(\left[\alpha_\mu,\alpha_\nu\right]
\left[\alpha^\mu,\alpha^\nu\right]\right)\nonumber\\
&&\hspace{0.1cm}
+{\rm tr}\
\left\{{\cal M}\left[\beta^\prime\left(
\alpha_\mu \alpha_\mu U
+U^{\dag}\alpha_\mu \alpha_\mu\right)
+\delta^\prime\left(U+U^{\dag}-2\right)\right]\right\}~,
\ee
where 'tr' denotes the trace in flavor space, $f_\pi \simeq 93\,$MeV the
pion decay constant and $e$ is a dimensionless number (the so--called
Skyrme parameter), which can be {determined}
from a best fit to baryon observables. Typically, $e\simeq 4$.
The chiral field $U$, which parametrizes the Goldstone boson
octet, is contained in
$\alpha_\mu=\partial_\mu U U^\dagger$ and
$\beta_\mu= U^\dagger \partial_\mu U$.
It has been demonstrated before that for realistic three flavor
calculations, one has to include symmetry breaking terms of the
form given in eq.(\ref{eq:LS}).
The parameters associated with the symmetry breaking
are determined via
\be
m_\pi^2=\frac{4\delta^\prime}{f_\pi^2}\ , \quad
m_K^2=\frac{2(1+x)\delta^\prime}{f_K^2}\ \quad {\rm and}\quad
\left(\frac{f_K}{f_\pi}\right)^2=
1+4\beta^\prime\frac{1-x}{f_\pi^2},
\ee
as ${\cal M}=y\lambda^3+T+xS$ is proportional to the current quark mass
matrix (for a detailed discussion, see ref.~\cite{HW}), $m_\pi$ ($m_K$)
is the pion (kaon) mass and $f_K = 1.2f_\pi$ the kaon decay constant.
Here we ignore isospin breaking, {\it i.e.} we set $y=0$.
Finally the Wess--Zumino--Witten term has to be added to the action
\be
\Gamma_{\rm WZ}=-\frac{iN_C}{240\pi^2}
\int_{M_5}d^5x\, \epsilon_{\mu\nu\rho\sigma\tau}\,
\alpha^\mu\alpha^\nu\alpha^\rho\alpha^\sigma\alpha^\tau \ ,
\ee
with $N_C$ the number of colors.
The classical soliton is obtained from the hedgehog {\it ansatz}
\be
U_0(\mbox{\boldmath $r$})=\pmatrix{
{\rm exp}\left(i\mbox{\boldmath $\tau$}\cdot
\hat{\mbox{\boldmath $r$}}F(r)\right) &
{| \atop |}&\hspace{-10pt}{\mbox{\small $0$}
\atop \mbox{\small $0$}}\cr
-------\ -\hspace{-8pt}&-&\hspace{-10pt}-----\cr
0\qquad 0&|&\hspace{-8pt} 1\cr}.
\ee
Here, $F(r)$ is the so--called chiral angle which is obtained from
minimization of the soliton mass. Since the classical soliton has neither
good isospin (hypercharge) nor spin angular momentum, one has to perform
an adiabatic rotation to obtain the physical baryon states.
The pertinent collective coordinates are introduced via {\cite{WSPM}}
\be
U(\mbox{\boldmath $r$},t)= A^{\dagger}(t)\
{\rm e}^{iZ(\mbox{\scriptsize\boldmath $r$})} \sqrt{U_0} \
{\rm e}^{2iZ(\mbox{\scriptsize\boldmath $r$})}
\sqrt{U_0} \ {\rm e}^{iZ(\mbox{\scriptsize\boldmath $r$})}A(t)\ ,
\quad
Z(\mbox{\boldmath $r$})=\frac{1}{2}\pmatrix{\mbox{\large $0$} &
\hspace{-10pt}{| \atop |} &
\hspace{-10pt}K(\mbox{\boldmath $r$}) \cr
---\hspace{-12pt}&-&\hspace{-10pt}---\cr
\ K^{\dagger}(\mbox{\boldmath $r$})&
\hspace{-10pt} |&\hspace{-8pt} 0\cr},
\ee
and the time dependence of the collective coordinates $A$ is measured
by the angular velocities $\Omega^a$:
\be
A^\dagger\frac{d}{dt} A=\frac{i}{2}\sum_{a=1}^8 \lambda^a\Omega^a .
\ee
Kaon fields $K(\mbox{\boldmath $r$})$ are induced by a linear 
coupling to the angluar velocities.
This linear coupling term is contained in the Wess--Zumino action. An
appropriate {\it ansatz} to the corresponding variational equations is
\be
K(\mbox{\boldmath $r$})=W(r)\mbox{\boldmath $\tau$}\cdot
\hat{\mbox{\boldmath $r$}}\Omega_K\ , \quad
\Omega_K=\frac{1}{2}
\pmatrix{\Omega_4-i\Omega_5 \cr\Omega_6-i\Omega_7}.
\ee
These induced fields have a nonvanishing overlap
\be
\langle K | K_0 \rangle \propto
\int_0^\infty r^2 dr\, W(r)\, {\cal G}\,
\frac{{\rm sin}\frac{F}{2}}
{1-{\rm cos}\frac{F}{2}}
\ee
with the would--be zero--mode
$W_0(r)=(1-{\rm cos}\frac{F}{2})/{\rm sin}\frac{F}{2}$. Demanding
that there is no double counting of $W_0$ for the strange
moment of inertia requires the constraint
$\langle K | K_0 \rangle \equiv 0$ with the metric {\cite{Hans}}
\be \label{metric}
{\cal G}=\left\{2f_\pi^2+\frac{1}{2e^2}
\left(F^{\prime2}r^2+2{\rm sin}^2F\right)\right\}
\left(1-{\rm cos}F\right)\ .
\ee
However, we will consider various forms for ${\cal G}$ to illuminate
the effects of the induced kaon fields.
One finally ends up with the Hamiltonian for the
collective coordinates
\be
H=M_{\rm cl}+\frac{1}{2\alpha^2}\, J(J+1)
+\frac{1}{2\beta^2}\left[C_2-J(J+1)-\frac{3}{4}\right]
+\frac{1}{2}\gamma\left(1-D_{88}\right)
\ee
The classical mass $M_{\rm cl}$, the moment of inertia $\alpha^2$
and the symmetry breaker $\gamma$ are functionals of  the chiral
angle $F$ only. The induced kaon field $W(r)$ is obtained by
maximizing the strange moment of inertia $\beta^2$ subject to
the above constraint. $C_2=\sum_{a=1}^8 R_a R_a$ is the quadratic
Casimir operator defined in terms of the right {SU(3)}
generators $R_a$. The latter are the momentum operators conjugate
to the angular velocities,{\it i.e.} $R_a=-
\partial L/\partial \Omega_a$. Finally the adjoint representation
\be
D_{ab}=\frac{1}{2}{\rm tr}\left(\lambda_a A \lambda_b A^\dagger\right)
\ee
has been introduced. Exact diagonalization of $H$ yields the
wave functions of the low lying $\frac{1}{2}^+$ and $\frac{3}{2}^+$
baryons in the space of the collective coordinates.
Having defined the model, we proceed to calculate the pv pion--nucleon
coupling constant.

\medskip

\noindent {\bf 3.}
In the standard model the parity--violating interaction is contained in
the coupling of the heavy vector bosons ($W_\mu^\pm , Z_\mu$) to the
charged and neutral quark currents, denoted
$J^\mu_{\rm ch}$ and $J_{\rm n}^\mu$, respectively,
\be
\cL_{\rm PV}=\frac{g_2}{\sqrt{8}}\left( W_\mu^+ J^\mu_{\rm ch}
+W_\mu^-(J^\mu_{\rm ch})^\dagger\right)
-\frac{g_2}{2{\rm cos}\Theta_W}Z_\mu J_{\rm n}^\mu~.
\ee
Here, $\Theta_W$ is the weak mixing angle (Weinberg angle).
These currents are identified with combinations of
the vector and axial--vector currents {\cite{PSW}}
of the three flavor Skyrme model (for $N_C=3$)
(for more details, see e.g. ref.~\cite{KM2})
\be
V_\mu^a (A_\mu^a) & = & -\frac{i}{2}f_\pi^2\
{\rm tr}\left\{Q^a\left(\alpha_\mu\mp\beta_\mu\right)\right\}
-\frac{i}{8e^2}{\rm tr}\left\{Q^a\left(
\left[\alpha_\nu,\left[\alpha_\mu,\alpha_\nu\right]\right]\mp
\left[\beta_\nu,\left[\beta_\mu,\beta_\nu\right]\right]
\right)\right\}
\nonumber \\ &&
-\frac{1}{16\pi^2}\epsilon^{\mu\nu\rho\sigma}
{\rm tr}\left\{Q^a\left(\alpha_\nu\alpha_\rho\alpha_\sigma
\pm\beta_\nu\beta_\rho\beta_\sigma\right)\right\}
\nonumber \\ &&
-i\beta^\prime{\rm tr}\left\{Q^a\left(
\{{U{\cal M}+\cal M}U^{\dag},\alpha_\mu\}\mp
\{{\cal M}U+U^{\dag}{\cal M},\beta_\mu\}\right)\right\},
\ee
where the $Q^a=(\frac{1}{3},\frac{\lambda^1}{2},\ldots,
\frac{\lambda^8}{2})$ denote the Hermitian nonet generators.
Elimination of the gauge bosons in the small momentum limit
($g_2^2/4m_W^2=\sqrt{2}G_F$)
yields the current--current interaction~\cite{KM1}
\beq
\cL_{\rm PV} = \sqrt{2}G_F\left\{
\sum_{i=1}^2 V_\mu^i A^{i\mu}
-\left[{\rm cos}(2\Theta_W)\left(V_\mu^3+\frac{1}{\sqrt{3}}V_\mu^8\right)
-\frac{1}{2}V_\mu^0\right]
\left[\frac{1}{2}V^{0\mu}-A^{3\mu}-\frac{1}{\sqrt{3}}A^{8\mu}\right]
\right\}~.
\eeq
Note that with $J_\mu^0 \to (2/3) J_\mu^0$ and $J_\mu^8 \to J_\mu^0
/ \sqrt{3}$ ($J_\mu = V_\mu , A_\mu$) one recovers the two--flavor
result of ref.~\cite{KM2}. As noted before, for the pure SU(2)
Skyrmion, the symmetry between the axial, vector and isoscalar
currents leads to a vanishing $G_\pi$. This accidental symmetry is
broken when one includes vector mesons or works in a realistic three
flavor version as done here. Furthermore, as argued in ref.~\cite{KM1}, the
so--called strong interaction enhancement factors are contained in the
non--perturbative soliton model currents.

\smallskip

\noindent
The pv pion--nucleon coupling constant $G_\pi$ is defined via
\be
\cL^{\rm PV}_{\pi N}= -\sqrt{\frac{1}{2}} \, G_\pi \, \bar{\Psi}_N \,
( \mbox {\boldmath $\tau$} \times
\mbox {\boldmath $\pi$}  \, )_3 \, \Psi_N~,
\ee
where $\Psi_N$ denotes the nucleon spinor and
$\mbox{\boldmath  $\tau$}$ the conventional
Pauli isospin matrices. From the above current--current interaction, we
have to extract the terms linear in the pion field $\pi$. This is done
by considering pionic fluactuations around the chiral field $U$ (see
e.g. ref.~\cite{KM2})
\be
U\rightarrow \eta \,  U \,  \eta \qquad , \qquad
\eta={\rm exp}\left(\frac{i}{2f_\pi}
\mbox{\boldmath $\tau$}\cdot \mbox{\boldmath $\pi$}\right)
\approx 1+\frac{i}{2f_\pi}
\mbox{\boldmath $\tau$}\cdot \mbox{\boldmath $\pi$}
\ee
yielding the Kroll--Ruderman relations
\be
V_\mu^a\rightarrow V_\mu^a+\frac{1}{2f_\pi}F^{iab}\pi_i A_\mu^b
\,\, , \qquad
A_\mu^a\rightarrow A_\mu^a+\frac{1}{2f_\pi}F^{iab}\pi_i V_\mu^b
\ee
up to minor contributions form the kinetic symmetry breaking term
($\beta^\prime$, $\beta^{\prime\prime}$).\footnote{These small corrections
could be taken  into account but play no role for the later results.}
The interesting structure constants for the problem under consideration
are $F^{i8b}=F^{i0b}=0$ and $F^{ijb}=2\epsilon_{ijb}$.
The coupling constant $G_\pi$ is now obtained from the
matrix element of the term linear in $\pi$  in $\cL_{\rm PV}$
between nucleon states (with spin up):
\be
G_\pi&=&
\frac{G_F}{f_\pi}\, \langle p \uparrow|\int d^3r \Bigg\{
2\, {\rm sin}^2\Theta_W\left(A_\mu^3A^{\mu +}+V_\mu^3V^{\mu +}\right)
+\frac{1}{2}V_\mu^0V^{+\mu} \nonumber  \\
&&\hspace{2cm}
-{\rm cos}(2\Theta_W)\left[A_\mu^+
\left(\frac{1}{\sqrt{3}}A^{8\mu}-\frac{1}{2}A^{0\mu}\right)
+\frac{1}{\sqrt{3}}V_\mu^8V^{+\mu}\right] \Bigg\}|n\uparrow\rangle .
\ee
\noindent
Here the superscript '$^+$' denotes the standard spherical
component $V^+=(V_1+iV_2)/\sqrt2$ and similarly
$A^+=(A_1+iA_2)/\sqrt2$.
The above matrix element is understood to be taken in the space of the
collective coordinates $A$ as the nucleon states are eigenstates
of the collective Hamiltonian $H$. The relevant operators are obtained
by subsituting the meson fields in the covariant expressions for the
currents. To be precise ($i,j,k=1,2,3$; $\alpha,\beta=4,..,7$)
\be
V^a_i&=&V_1(r)\eijk x_j D_{ak}+
\frac{\sqrt{3}}{2\alpha^2}B(r)\eijk x_j R_k D_{a8}
-\frac{1}{\beta^2}V_2(r)\eijk x_j d_{k\alpha\beta}D_{a\alpha}R_\beta
\nonumber \\ &&\hspace{1cm}
+V_3(r)\eijk x_j D_{88}D_{ak}+V_4(r)\eijk x_j
d_{k\alpha\beta}D_{a\alpha}D_{8\beta} \quad (a=1,..,8)
\ee
\be
V^a_0=\frac{\sqrt{3}}{2\alpha^2}B(r)D_{a8}
+\frac{1}{\alpha^2} V_7(r) D_{ai}R_i
+\frac{1}{\beta^2}V_8(r) D_{a\alpha}R_\alpha
\ee
\be
V_0^0=B(r)\, ,\qquad V_i^0=\frac{1}{\alpha^2}B(r)\eijk x_j R_k\, .
\ee
for the vector current and
\be
A_i^a&=&\left[A_1(r)\delta_{ij}+A_2(r){\hat x}_i{\hat x}_j\right] D_{aj}
-\frac{1}{\beta^2}\left[A_3(r)\delta_{ij}+A_4(r){\hat x}_i{\hat x}_j\right]
d_{j\alpha\beta}D_{a\alpha}R_\beta
\nonumber \\ && \hspace{1cm}
+\left[A_5(r)\delta_{ij}+A_6(r){\hat x}_i{\hat x}_j\right] D_{aj}D_{88}
+\left[A_7(r)\delta_{ij}+A_8(r){\hat x}_i{\hat x}_j\right] D_{a8}D_{8j}
\nonumber \\ && \hspace{1cm}
+\left[A_9(r)\delta_{ij}+A_{10}(r){\hat x}_i{\hat x}_j\right]
d_{j\alpha\beta}D_{a\alpha}D_{8\beta}
\label{eq22}
\ee
for the octet axial vector current. In the latter case the time
component vanishes while the axial singlet current is omitted as
it only acquires a negligibly small contribution from the
$\beta^\prime$ type symmetry breaker.  In the above equations the
angular velocities have been eliminated in favor of the SU(3)
generators  $R_a$. The radial functions
$V_1, B,..,A_1,...,A_{10}$ can be extracted from the literature
(see e.g. refs.~\cite{PSW},\cite{PW}).

\noindent
The collective matrix elements are evaluated by insertion of a
complete set of states, for example
\be
\langle p \uparrow|A_\mu^+ A^{8\mu} |n\uparrow\rangle=
\sum_{B;I,J}\langle p \uparrow|A_\mu^+|
B;I,J\rangle\langle B;I,J |A^{8\mu} |n\uparrow\rangle
\ee
with the sum running over all baryon eigenstates $B$ (with isospin $I$
and spin $J$) of the collective Hamiltonian $H$ to which the nucleon
can couple. In the absense of flavor symmetry breaking these would
be nucleon states in higher dimensional SU(3) representations. With
the inclusion of the $\gamma$ term in the Hamiltonian
$H$ these intermediate states are distorted as well.

\medskip

\begin{figure}[htb]
   \vspace{0.5cm}
   \epsfysize=10cm
   \centerline{\epsffile{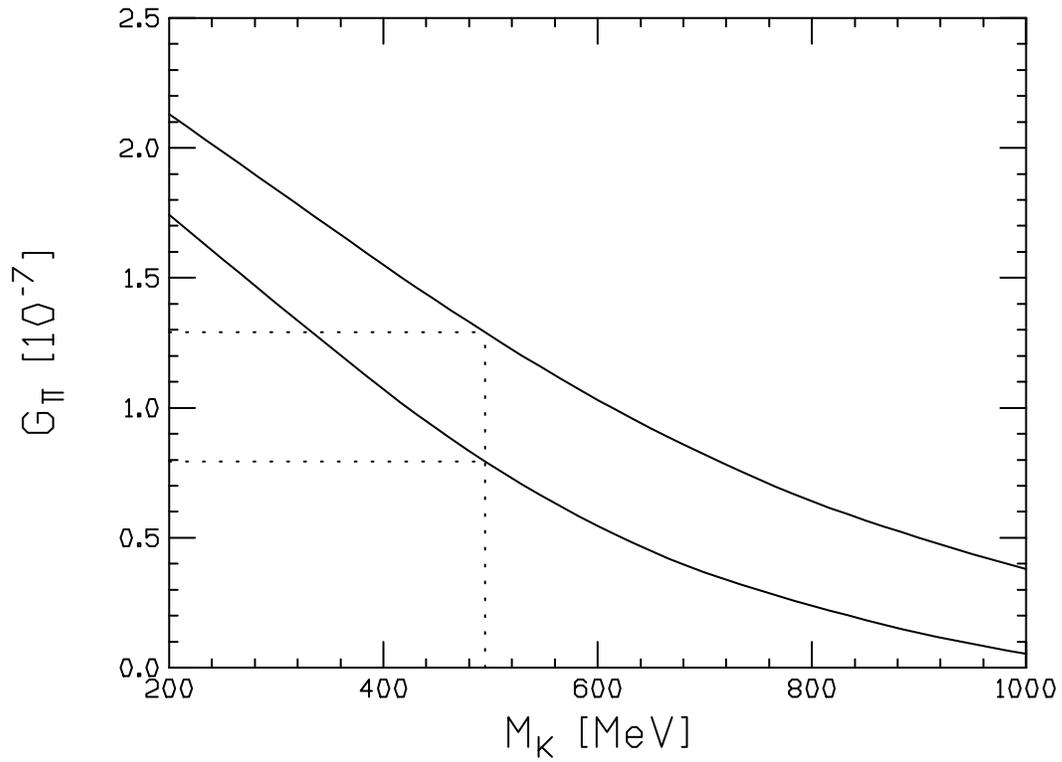}}
   \vspace{2.2cm}
   \centerline{\parbox{15cm}{\caption{\label{fig1}
   The weak pion--nucleon coupling constant as a function of the
   kaon mass. Upper (lower) solid curve: $e = 4.5 \, (4.0)$.
   For $m_K \to \infty$, one recovers the  SU(2) result.
   The dotted lines refer to the physical kaon mass.
}}}
\end{figure}

\begin{figure}[htb]
   \vspace{0.5cm}
   \epsfysize=12cm
   \centerline{\epsffile{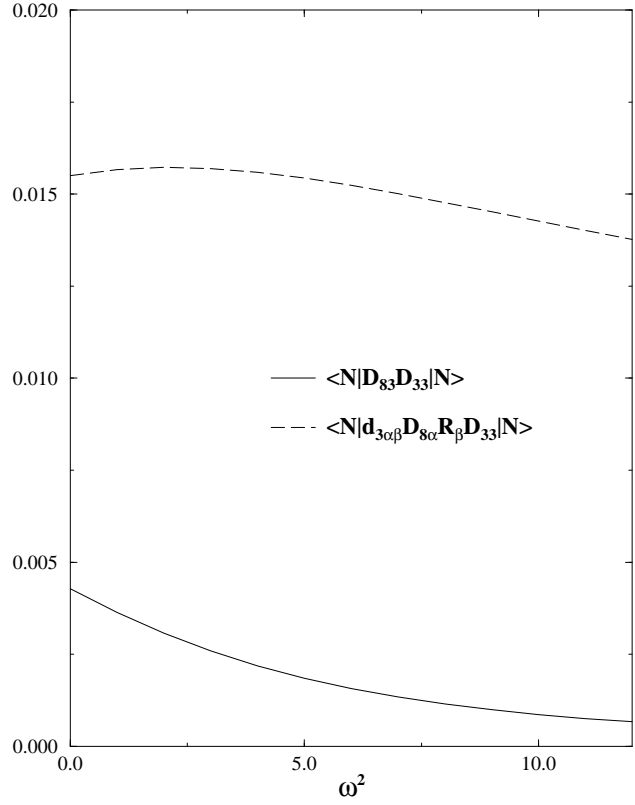}}
   \vspace{2.2cm}
   \centerline{\parbox{15cm}{\caption{\label{fig2}
   SU(3) matrix elements as a function of the symmetry breaking.
   The solid (dashed) line refers to the classical (induced) kaon
   fields. For realistic symmetry breaking,
   $\omega^2=\frac{3}{2}\gamma\beta^2 \simeq 5$.
}}}
\end{figure}

\noindent {\bf 4.}
We are now in the position to present the results. As input, we use
$f_\pi = 93\,$MeV, $f_K = 113\,$MeV, $m_\pi = 138\,$MeV,
$m_K = 495\,$MeV, $G_F = 1.16\cdot 10^{-5}\,$GeV$^{-2}$ and $\sin^2
\Theta_W = 0.23$. The Skyrme parameter is varied in the range
$e = 4.0 \ldots 4.5$, as our central value we use $e = 4.25$
which reasonably reproduces the baryon spectrum.
The results for a large number of hyperon properties for these parameters
can be found in the literature~\cite{HW}.
In fig.~\ref{fig1} we show the weak pion--nucleon coupling constant as a
function
of the kaon mass, i.e. as a function of the symmetry breaking. For the
physical value of $m_K$, we get
\beq\label{Gpival}
G_\pi = \{0.8 , 1.3\} \cdot 10^{-7} \quad {\rm for} \quad e = \{4.0,4.5\}~,
\eeq
which is considerably bigger than the SU(2) generalized Skyrme result
of $0.2 \ldots 0.3 \cdot
10^{-7}$~\cite{KM1,KM2}.\footnote{Remember that in the SU(2) Skyrme
  model without vector mesons, one has $G_\pi = 0$~\cite{KM1}.}
However, as one freezes out the kaon degrees
of freedom, the value of $G_\pi$ approaches zero as
indicated in fig.~\ref{fig1}. The values given in eq.(\ref{Gpival})
are in fair agreement with most recent quark model calculations in which
$G_\pi = 2.0 \ldots 2.7\cdot 10^{-7}$~\cite{BD,FCDH}. The typical
range of values in the quark model calculations is $G_\pi = 0 \ldots 3
\cdot 10^{-7}$~\cite{BD}. This large enhancement of
the weak pion--nucleon coupling compared to the SU(2) calculations
is largely due to the induced kaon
fields as shown in fig.~\ref{fig2}. Given  in that figure are the
proton matrix elements of the operators
$O_1 = d_{3\alpha\beta}D_{8\alpha}R_\beta D_{33}$
and $O_0=D_{83}D_{33}$.  Assuming isospin invariance these operators
appear, for example, when evaluating the matrix element
$\langle p | A_i^+A_i^8 | n \rangle$, {\it cf.} eq. (\ref{eq22}).
While the coefficient of the former contains
the induced kaon fields, the latter one is given entirely in terms of
the classical fields.  A typical example for a coefficient of
$O_1$ would be $A_1\times A_3$ while the one of $O_0$ is just
like $(A_1)^2$. Again,
for the explicit expressions of these radial functions we
refer to the literature \cite{PSW}.
We note that the matrix element of $O_1$ is not
only sizeably bigger than that of $O_0$ but also it is almost constant
under symmetry breaking while $O_0$ has dropped to half of its
flavor symmetric value at $\omega^2 =3\gamma\beta^2 /2 = 0$
for realistic symmetry
breaking $\omega^2\approx 5$.  Actually the matrix element of
$O_1$  has a positive slope at $\omega^2=0$. This underlines
the importance of the induced components for the observable under
consideration. These results look at first glance surprising since
for most observables, the induced kaon fields play only a minor
role~\cite{HW}. However, a closer look at table~1 of ref.~\cite{PSW}
reveals that the isoscalar magnetic moments ($n+p, \Lambda, \Sigma^0$)
are given up to 50\% from the induced components (the $V_2$
contribution in that table). Furthermore, the isoscalar part of $V_2$ contains
matrix elements of the operator $d_{3\alpha\beta}D_{8\alpha}R_\beta$.
These contributions are not SU(3) symmetric by themselves but in the
case of the magnetic moments tend to restore the SU(3)
symmetry~\cite{PSW}. For the weak pion--nucleon coupling constant,
this means that four--fermion operators of the type $(\bar q q) (\bar
s s)$ contribute significantly  because they are enhanced by the
induced kaon fields. More quantitatively, setting $W(r) = 0$ reduces
the value of $G_\pi$ from $1.1 \cdot 10^{-7}$ to $0.5 \cdot 10^{-7}$
for $e=4.25$. This has to be contrasted to the small
induced kaon contribution to the strange moment
of inertia $\beta^2$. Using $e=4.25$ we find $\beta^2=0.27/{\rm GeV}$ setting
$W(r)$ equal to zero this is only moderately reduced to $0.24/{\rm GeV}$.
This underlines the statement that for three
flavor calculations, one has to work with realistic symmetry breaking.
Formally, such effects are suppressed in the large color limit
but for the real world with $N_C =3$ they have to be taken into
account as clearly demonstrated in this calculation. We stress that
the large enhancement of $G_\pi$ compared to the two--flavor case is
not an effect of a large strangeness component in the nucleons' wave
function, but rather the suppression of the four--fermion operators
involving strange quark pairs in the two--flavor case.
We have furthermore studied the dependence on the metric ${\cal G}$. For
$e=4.25$ and $m_K = 495\,$MeV, setting ${\cal G} = 0 \,
(1-{\rm cos}F)$ leads to
$G_\pi = 2.6 \, (1.9) \cdot 10^{-7}$, and the induced kaon fields
contribute as much as 40\% (25\%) to $\beta^2$. It is therefore
mandatory to account for the correct metric ${\cal G}$ as given in
eq.(\ref{metric}).
Let us finally also note that similarly to $G_\pi$ one can
evaluate the widths of non--leptonic hyperon decays in the
standard model starting from the current--current interaction.
In a recent study \cite{NNS} it has been observed that also for these
observables an important enhancement is due to the inclusion of
strange degrees of freedom.

\medskip

\noindent {\bf 5.} In summary, we have presented the results of a
three flavor Skyrme model calculation for the parity--violating pion
nucleon coupling constant. The resulting number of $G_\pi = 0.8 \ldots
1.3 \cdot 10^{-7}$ is in fair agreement with most recent quark model
calculations and is
considerably bigger than previously found values in two--flavor Skyrme
type models. This large enhancement underlines the importance of
four--fermion operators involving strange quark pairs. For the
parity--violating vector meson couplings to the nucleon, one does not
expect such dramatic changes compared to the two--flavor calculation
since no unnatural suppression is involved in these
cases. Nevertheless, a detailed study in a realistic three flavor
vector meson Skyrme model should be  carried out. In addition, one
can also investigate the parity--violating vertices including the
$\Delta$ resonance and the extension of the pertinent interaction
regions.

\bigskip
\bigskip
\bigskip
\bigskip
\bigskip

\noindent {\bf Acknowledgements}

\medskip

\noindent We are grateful to Mike Musolf and Martin Savage for some
pertinent remarks. One of us (UGM) thanks the Institute for Nuclear
Theory at the University of Washington for its hospitality and the
DOE for partial support during the completion of this work.

\newpage

\noindent {\large {\bf Figures}}

\bigskip
\bigskip
\bigskip
\bigskip
\bigskip
\bigskip
\bigskip
\bigskip
\bigskip
\bigskip

\bigskip
\bigskip
\bigskip
\bigskip

\end{document}